\documentclass[twocolumn]{autart} 
\usepackage{amssymb}
\usepackage{color}
\usepackage{graphicx}
\usepackage{amssymb}
\usepackage{hyperref}
\hypersetup{
  colorlinks   = true, %Colours links instead of ugly boxes
  urlcolor     = blue, %Colour for external hyperlinks
  linkcolor    = blue, %Colour of internal links
  citecolor   = red %Colour of citations
}
\usepackage{epstopdf}
\usepackage{amsfonts}
\usepackage{mathrsfs} 
\usepackage{psfrag}
\usepackage{chngcntr}
\usepackage{apptools}
\AtAppendix{\counterwithin{theorem}{section}}
\AtAppendix{\counterwithin{property}{section}}
\AtAppendix{\counterwithin{lemma}{section}}
\usepackage[crop=pdfcrop]{pstool}
\usepackage{amssymb}
\usepackage{amsmath} % assumes amsmath package installed 
\usepackage{mathtools}
\usepackage{balance}
\usepackage{algorithm}% http://ctan.org/pkg/algorithms
\usepackage{algpseudocode}% http://ctan.org/pkg/algorithmicx
\usepackage{tikz}
\usetikzlibrary{shapes,arrows}
\newtheorem{remark}{Remark}
\newtheorem{assumption}{Assumption}
\newtheorem{property}{Property}

\newtheorem{lemma}{Lemma}
\newtheorem{proposition}{Proposition}
\newtheorem{problem}{Problem}

\newtheorem{definition}{Definition}

\newcommand{\Gt}{\mathcal{\widehat{G}}}

\newcommand{\gxc}{h}
\newcommand{\sats}{\operatorname{sat}}
\newcommand{\sat}{\mathfrak{sat}_{\bar{u}}}
\newcommand{\dz}{\mathfrak{dz}_{\bar{u}}}
\newcommand{\id}{\operatorname{id}}
\newcommand{\gph}{\operatorname{gph}}
\newcommand{\D}{\mathcal{D}}
\newcommand{\Dy}{\mathbb{D}}
\newcommand{\C}{\mathcal{C}}
\newcommand{\U}{\mathcal{U}}
\renewcommand{\S}{\mathbb{S}}
\newcommand*{\QEDB}{\hfill\ensuremath{\square}}%
\newcommand*{\QEDA}{\hfill\ensuremath{\blacksquare}}%
\newcommand{\A}{\mathcal{A}}
\newcommand{\T}{\mathcal{T}}
\newcommand{\R}{\mathbb{R}}
\newcommand{\Nats}{\mathbb{N}}
\newcommand{\Hcl}{\mathcal{H}_{cl}}
\newcommand{\nuu}{m}
\newcommand{\np}{n}
\newcommand{\ns}{q}
\newcommand{\Rp}{\mathbb{R}^{n}}
\newcommand{\Rc}{\R^{\nuu}}

\newcommand{\Spn}{\mathbb{S}^{n}_{+}}
\newcommand{\xp}{x_p}
\newcommand{\xpi}{x^{(i)}_p}
\newcommand{\xci}{\eta^{(i)}}
\newcommand{\xcl}{\bar{x}}
\newcommand{\dotxp}{\dot{x}_p}
\newcommand{\N}{\mathcal{N}}
\newcommand{\nx}{2n+q}
\newcommand{\xc}{\eta}
\newcommand{\tildexci}{\tilde{\eta}^{(i)}}
\newcommand{\dotxc}{\dot{\xc}}

\newcommand{\Q}{\mathcal{N}}

\newcommand{\dom}{\operatorname{dom}}
\newcommand{\Sign}{\operatorname{sign}}
\newcommand{\Id}{\mathbf{I}}
\newcommand{\xctilde}{\tilde{\eta}}
\newcommand{\minimize}{\operatorname{minimize}}
\newcommand{\Tr}{\operatorname{trace}}
\newcommand{\range}{\operatorname{rge}}
\newcommand{\Co}{\operatorname{co}}
\newcommand{\He}{\operatorname{He}}

\newcommand{\0}{{\bf 0}}
\newcommand{\1}{{\bf 1}}

\newcommand*{\tr}{^{\mkern-1.5mu\mathsf{T}}}

\newcommand{\ie}{{i.e.}}

\newtheorem{theorem}{Theorem}
\begin{document}
\begin{frontmatter}
\title{\LARGE \bf Control Design under Actuator Saturation and Multi-Rate Sampling\\ (Extended Version)}
\thanks[footnoteinfo]{Research by R. G. Sanfelice partially supported by NSF Grants no. ECS-1710621, CNS-2039054, and CNS-2111688, by AFOSR Grants no. FA9550-19-1-0053, FA9550-19-1-0169, and FA9550-20-1-0238, and by ARO Grant no. W911NF-20-1-0253. Research by Francesco Ferrante and Sophie Tarbouriech is funded in part by ANR via project HANDY, number ANR-18-CE40-0010.}
\author[GR]{Francesco Ferrante},
\ead{francesco.ferrante@unipg.it}
\author[UCSC]{Ricardo G. Sanfelice},
\ead{ricardo@ucsc.edu}
\author[LAAS]{Sophie Tarbouriech}
\ead{tarbour@laas.fr}
\address[GR]{Department of Engineering, University of Perugia, Via G. Duranti, 67, 06125 Perugia, Italy.}
\address[UCSC]{Electrical and Computer Engineering Department, University of California,
Santa Cruz, CA 95064, USA.}
\address[LAAS]{LAAS-CNRS, Universit\'e de Toulouse, CNRS, Toulouse, France.}
\begin{abstract}
The problem of designing a stabilizing feedback controller in the presence of 
saturating actuators and multi-rate (asynchronous) aperiodic state measurements is studied. Specifically, we consider a scenario in which measurements of the plant states are collected at the controller end in a sporadic and asynchronous fashion. A hybrid controller is used to perform a fusion of measurements sampled at different times. In between sampling events, the controller behaves as a copy of the plant and provides a feedback control signal based on the reconstruction of the plant state. The presence of saturation at the plant input limits the value of the components of this signal to a bounded range.
When a new measurement is available, the controller state undergoes an instantaneous jump. The resulting system is augmented with a set of timers triggering the arrival of new measurements and analyzed in a hybrid systems framework. Relying on 
Lyapunov tools for hybrid systems and techniques for control design under saturation, we propose sufficient conditions in the form of matrix inequalities to ensure regional exponential 
stability of a closed-set containing the origin of the plant, i.e., exponential stability with a guaranteed region of attraction. 
Specifically, explicit estimates of the basin of attraction are provided in the form of ellipsoidal sets.
Leveraging those conditions, a design procedure based on semidefinite programming is proposed to design a stabilizing controller with maximized size of the basin attraction.  
%Furthermore, robustness of the closed-loop system to measurement noise and small general perturbations is studied. 
The effectiveness of the proposed methodology is shown in an example.
\end{abstract}
\end{frontmatter}
\section{Introduction}
%\subsection{Motivation and Background}
Control of sampled-data systems has been attracting the attention of researchers for a long time and its interest is still vivid in the community as pointed out by the survey \cite{hetel2017recent}. Indeed, capturing the effect of sampled-data information in control loops is fundamental in addressing the challenges posed by modern engineered systems, in which the use of embedded systems and communication networks is central; see, e.g., \cite{hespanha2007survey} and the references therein.
%In this paper, we are interested in the design of robust stabilizing controllers in the presence of actuator saturation and asynchronously aperiodically sampled state measurements. The main difficulties in the solution to this problem can be summarized as follows:
%\begin{itemize}
%\item \textit{Asynchronously aperiodically sampled measurements}: measurements are  gathered at different times and the time in between consecutive measurements may be time-varying.
%\item \textit{Actuator saturation}: global asymptotic stabilization 
%may be impossible to achieve with bounded inputs; see, e.g., \cite{Sontag1989}. In this 
%case, the basin of attraction of the closed-loop system in bounded and is hard to 
%characterize.
%\item \textit{Uncertainties}: Plant dynamics may be uncertain and large parameter 
%uncertainties may induce instability and poor convergence properties. 
%\end{itemize} 
%\subsection{Related Work}
One of the most challenging aspects in the literature of sampled-data systems pertains to the study of aperiodic sampled-data systems, that is systems in which the exchange of information from the plant to the controller happens in a sampled-data aperiodic fashion. In fact, aperiodic sampling may be used as a modeling paradigm to capture different phenomena occurring in data networks such as dropouts, sampling jitter, etc; see \cite{hetel2017recent} for a survey on sampled-data systems. Essentially three main methodologies can be found in the literature for the analysis of sampled-data systems. The first consists of a time-varying discrete-time representation of the sampled-data system. This approach is pursued, e.g., in \cite{oishi2010stability,seuret2012novel}.  The second category of methodologies builds upon a time-delay model of the sampling mechanism and the use of Lyapunov-Krasovskii functionals for stability analysis. Following this approach, sufficient conditions for closed-loop asymptotic stability of linear aperiodic sampled-data systems are proposed in \cite{fridman2010refined,fujioka2009stability}. 
The third group of methodologies originates from the use of hybrid/impulsive models. This line of research is explored in \cite{naghshtabrizi2008exponential}, where Lyapunov functions with discontinuity at the sampling times are exploited to derive sufficient conditions for stability of linear aperiodic sampled-data systems. Computationally tractable conditions for stability analysis of linear aperiodic impulsive systems are given in \cite{briat2013convex}. More general stability results for nonlinear sampled-data systems based on the hybrid systems framework in \cite{goebel2012hybrid} are given in \cite{nesic2009explicit}.

A common feature in the above references consists of assuming synchronous sampling of different variables. However, in certain applications, due to technological constraints, some variables may happen to be measured at different sampling rates than others, making the assumption of synchronous sampling restrictive; see, e.g., \cite{antunes2010design,nagamune2005multirate}.
Because of its practical relevance, this problem has been attracting the attention of researchers for a long time \cite{kranc1957input,colaneri1990stabilization} and  \cite{lall2001lmi,naghshtabrizi2008exponential,moarref2014stability} for more recent contributions.
%Using a lifted representations and impulsive models of multi-rate sample-data systems, in \cite{lall2001lmi} the authors establish necessary and sufficient conditions for $\mathcal{H}_\infty$ control for linear systems in which inputs and outputs are updated, respectively, sampled using  asynchronous aperiodic sampling rates. The approach therein consists of recasting  controller design for multi-rate sampled-data systems as a convex optimization problem over linear operator inequalities. However, these linear operator inequalities are difficult to exploit for controller design and can be effectively adopted only in the presence of periodic commensurate sampling. To overcome this drawback, building upon the general framework in \cite{naghshtabrizi2008exponential}, in \cite{moarref2014stability} the authors establish sufficient conditions for stability analysis of multi-rate sampled-data systems with nonuniform sampling and input update and incommensurable sampling rates. 
%
A common characteristic in all the works cited so far consists of adopting zero-order-holder devices to generate intersample control inputs, i.e., the control input is held constant between sampling events. More general ``holding functions'' are proposed in \cite{garcia2013model} for multi-rate discrete-time systems. In particular, in \cite{ahmed2016} the authors show how exponentially decaying holding functions can be used to tolerate larger sampling rates in linear single-rate sampled-data systems.

Another limitation omnipresent in control systems pertains to actuator saturation. Due to its practical relevance, the interplay of sampling and saturation has attracted the interest of researchers over the last decade. Synchronous sampled-data $\mathcal{H}_\infty$ control design for linear plants subject to input saturation is addressed in \cite{dai2010output} via hybrid systems tools. Stability analysis and control design for linear systems under aperiodic saturated state/output feedback are tackled in \cite{fridman2004robust}, and later refined in 
\cite{da2016regional,seuret2012taking}, where the looped-functional approach in \cite{briat2012looped,seuret2012novel} is adopted.

Although the analysis of saturation in (single rate) sampled-data systems has received a lot of attention, despite the relevance of asynchronous sample-data systems in practical applications, to the best of our knowledge, no results are found in the literature to deal with actuator saturation in an asynchronous sampling setting. The objective of this paper is to fill this gap. To this end, we propose a controller that achieves local exponential stability by using information \emph{sampled in a sporadic multi-rate (asynchronous) fashion}. Our main contribution consists of sufficient conditions for the design of such a controller in a systematic way, i.e., via the solution to suitable optimization problems. 
As opposed to the case of networked control systems \cite{carnevale2007lyapunov} and event-triggered control systems \cite{forni2014event}, we assume that the communication between plant and controller is not ruled by a ``stabilizing" protocol and that only some upper bounds on the largest sampling rates are available. This is typically the case when variables are measured via sensors working at different sampling-rates or in the presence of network unreliabilities in transmission links. Our approach in solving this problem can be summarized as follows:
\begin{itemize}
\item We propose a hybrid controller that merges measurements sampled at different times and that provides a feedback control signal based on the reconstruction of the plant state. In between sampling events, the controller behaves as a copy of the plant, while when a new measurement is available, the controller state undergoes an instantaneous jump. This extends prior work in \cite{montestruque2004stability} to the case of input saturation.\\

%\item To capture the occurrence of asynchronous sampling events, we model the closed-loop system as a hybrid system with nonunique solutions.\\

\item Because of input saturation, achieving global exponential stability is in general impossible. To overcome this problem, we focus on regional exponential stability. In particular, as shown in Section~\ref{sec:Lyap}, by relying on the use of Lyapunov theory for hybrid systems in the framework \cite{goebel2012hybrid}, quadratic-like clock-dependent Lyapunov functions, and a characterization of the saturation based on regional sector conditions, we provide sufficient conditions in the form of matrix inequalities to assure closed-loop regional exponential stability. In this direction, one of the main contributions of this paper consists of the use of a clock-dependent Lyapunov function in a regional stabilization setting. The main difficulty to tackle in this case is that estimates of the basin of attraction (given in terms of sublevel set of the Lyapunov function) inherently depend on the ``artificial'' clock variable. To deal with this aspect, we adapt the local sector bound approach in \cite{tarbouriech2011stability}.\\

\item Building upon the results in Section~\ref{sec:Lyap}, Section~\ref{subsec:ControlDesign} proposes conditions that are specifically suited to controller design. One of the main contributions of this latter section is the formulation of a numerically tractable optimization problem for the design of a controller inducing basin of attraction maximization.\\

%\item Robustness of the closed-loop system is discussed in Section~\ref{sec:Rob}. Specifically, in Section~\ref{sec:RobNoise} we show that the closed-loop system is input-to-stable stable with respect to measurement noise. Moreover, in Section~\ref{sec:RobGen} we characterize robustness to small general perturbations such as uncertainties and variations of the bounds on the sampling rates and plant uncertainties and unmodeled dynamics.  
%\item The effect of large plant uncertainties is analyzed in Section~\ref{sec:Rob}. More precisely, in this section we provide sufficient conditions for the analysis of the impact of norm bounded plant uncertainties on the size of the basin of attraction, for which explicit estimates are given.
\end{itemize}
%The effectiveness of the proposed control design methodology is showcased in Section~\ref{sec:Ex} in a numerical example. 
It is worth to mention that the use of hybrid systems tools in the setup considered in this paper makes it possible to consider the effect of controller initial conditions on the closed-loop system. This is a relevant aspect in the case of saturating actuators since a mismatch in con- troller and plant initial conditions may have a dramatic impact on the behavior of the closed-loop system. 
In most of the results proposed in the literature, controller and plant initial conditions are assumed to be identical; see \cite{fridman2004robust,seuret2012novel} and the references therein, no assumption of this type is considered in our work.

This paper extends the results in our preliminary conference paper \cite{ferrante2018hybrid}. In particular, while in \cite{ferrante2018hybrid} we only focus on stability analysis, this work pertains to control design with basin of attraction maximization. In addition, here we provide full proofs of all the results; no proofs are included in the conference paper \cite{ferrante2018hybrid}.
\subsection{Notation}
The symbol $\Nats$ indicates the set of positive integers including zero, $\Nats_{>0}$ is the set of strictly positive integers, $\R_{\geq 0}$ represents the set of nonnegative real numbers, $\R^n$ is the Euclidean space of dimension $n$, $\R^{n\times m}$ represents the set of the $n\times m$ real matrices, $\R^n_{>0}$ is set of vectors in $\R^n$ with positive entries, $\S^{n}_{+}$ is the set of $n\times n$ symmetric positive definite matrices, and $\Dy^{n}_+$ denotes the set of $n\times n$ diagonal positive definite matrices. For a vector $x\in\R^n$, $\vert x \vert$ denotes the Euclidean norm and $x_i$ stands for the $i$-th component of $x$. Given vectors $x, y$, we use the equivalent notation $(x, y)=[x\tr\,\,y\tr]\tr$.
The identity matrix of size $n$ is denoted by $\Id_n$. The symbol $\1_{n}$ denotes the all-ones vector in $\R^n$. For a matrix $A\in\R^{n\times m}$, $A\tr$ denotes the transpose of $A$, $A_{(i)}$ the $i$-th row of $A$, $A\tr_{(i)}$ the $i$-th row of $A\tr$, and, when $n=m$, $\He (A)=A+A\tr$. For a symmetric matrix $A$, positive definiteness (negative definiteness) and positive semidefiniteness (negative semidefiniteness) are denoted, respectively, by $A\succ 0$ ($A\prec 0$) and $A\succeq 0$ ($A\preceq 0$). In partitioned symmetric matrices, the symbol $\star$ stands for symmetric blocks. Given matrices $A$ and $B$, $A\oplus B$ stands for the direct sum of $A$ and $B$, i.e., the block-diagonal matrix having $A$ and $B$ as diagonal blocks and $\displaystyle\bigoplus_{i=1}^n A_i$ is the direct sum of the matrices $A_1, A_2, \dots, A_n$. Given $x\in\R^{n}$ and a nonempty set $\mathcal{A}$, the distance of $x$ to $\mathcal{A}$ is defined as 
$\vert x \vert_{\mathcal{A}}=\inf_{y\in {\mathcal{A}}} \vert x-y \vert$. For any function $z\colon\R\rightarrow\R^n$, for any $t\in\R$, we denote $z(t^+)\coloneqq \lim_{s\rightarrow t^+} z(s)$. Given $P\in\Spn$ and $c>0$, $\mathcal{E}(P, c)\coloneqq\{x\in\R^n\colon x\tr Px\leq c\}$. Given a set $S$, we denote $\Co S$ the convex-hull of $S$. The symbol $\times$ denotes the Cartesian product of two sets, while given sets $S_1, S_2,\dots, S_n$, we use the notation $\bigtimes_{i=1}^n S_i=S_1 \times S_2\times\dots\times S_n$. The symbol $\range f$ stands for the image of the function $f$.
Let $c\in\R$, we denote by $L_V(c)$ the $c$-sublevel set of the function $V\colon \dom V\rightarrow\R$, i.e., $L_V(c)\coloneqq\{x\in\dom V\colon V(x)\leq c\}$. We denote by $V\vert_{K}$ the restriction of the funciton $V$ to the set $K$.
\subsection{Preliminaries on Hybrid Systems}
\label{sec:PreliminariesHybrid}
In this paper,  we consider hybrid systems with state $x\in\R^{n}$ of the form
\begin{equation}
\label{eq:Hybrid}
\mathcal{H}\left\lbrace
\begin{array}{ccll}
\dot{x}&=&f(x)&\quad x\in C\\
x^+&\in&G(x)&\quad x\in D
\end{array}\right.
\end{equation}
In particular we denote the function $f\colon\R^{n}\rightarrow\R^{n}$ as the \emph{flow map}, $C\subset\R^{n}$ as the \emph{flow set}, the set-valued map $G\colon\R^{n}\rightrightarrows\R^{n}$ as the \emph{jump map}, and $D\subset\R^{n}$ as the \emph{jump set}.
A set $E\subset\R_{\geq 0}\times \Nats$ is a \emph{hybrid time domain} if it is the union of a finite or infinite sequence of intervals $[t_j,t_{j+1}]\times\{j\}$, with the last interval (if existent) of the form $[t_j,T)$ with $T$ finite or $T=\infty$.  A function $\phi\colon \dom\phi\rightarrow\R^{n}$ is a hybrid arc if $\dom\phi$ is a hybrid-time domain and $\phi(\cdot, j)$ is locally absolutely continuous for each $j$. 
A hybrid arc $\phi$ is a solution to $\mathcal{H}$ if it satisfies the  dynamics of $\mathcal{H}$; see \cite{goebel2012hybrid}. A solution $\phi$ to $\mathcal{H}$ is maximal if it cannot be extended and is complete if $\dom\phi$ is unbounded; see~\cite{goebel2012hybrid} for more details on hybrid dynamical systems. 
\section{Problem Formulation}
\label{sec:ProbStat}
\subsection{System description}
Consider  the following continuous-time plant:
\begin{equation}
%\mathcal{P}
\left\lbrace
\begin{array}{ccll}
\dotxp(t)&=&A\xp(t)+Bu(t)\\
u(t)&=&\sat(v(t)),&
\end{array}\right.
\label{eq:Plant}
\end{equation}
where $\xp\in\Rp$ is the plant state, $u\in\Rc$ is the plant input and $v\in\Rc$ is the actuator input. Matrices $A\in\R^{\np\times\np}$ and $B\in\R^{\np\times \nuu}$ are given.
The function $v\mapsto \sat(v)$ is the symmetric decentralized saturation function with saturation level $\bar{u}=(\bar{u}_1, \bar{u}_2, \dots, \bar{u}_{\nuu})\in\R^{\nuu}_{>0}$. More precisely, 
\begin{equation}
v\mapsto \sat(v)\coloneqq (\sats_{1}(v_1), \sats_{2}(v_2), \dots, \sats_{\nuu}(v_{\nuu})),
\label{eq:sat}
\end{equation}
where for each $v_i\in\R$, $\sats_{i}(v_i)\coloneqq\min(\vert v_i \vert, \bar{u}_{i})\Sign(v_i)$, $i = 1, ..., \nuu$.

We assume that sporadic and asynchronous measurements of the components of the plant state $\xp$ are available. More precisely, assume that the components of the plant state $\xp$ are clustered into $\ns\leq \np$ subvectors $x^{(1)}_{p}\in\R^{n_1}, x^{(2)}_{p}\in\R^{n_2},\dots, x^{(\ns)}_{p}\in\R^{n_\ns}$, \ie, $\xp=(x^{(1)}_{p}, x^{(2)}_{p},\dots, x^{(\ns)}_{p})$  and that  measurements of $\xpi$ are  available only at some time instances $t^{(i)}_k$, $k\in\Nats$, not known {\itshape a priori}. 
For each $i\in\mathcal{N}\coloneqq\{1,2,\dots, \ns\}$, we define $M_i\subset\R_{\geq 0}$ as the set of measurement times of $\xpi$, \ie, $M_i=\{t_k^{(i)}\}_{k=0}^\infty$ and $\mathcal{M}=\bigcup_{i=1}^{\ns} M_i$ as the set of all measurement times. In this setting, our goal is to design a feedback controller ensuring regional closed-loop exponential stability of the origin of the plant \eqref{eq:Plant}, i.e., local exponential stability with an explicit estimate of the basin of attraction.

Inspired by \cite{montestruque2004stability},  we consider the following controller with  jumps in its state:
\begin{equation}
%\mathcal{K}
\left\lbrace
\begin{array}{ccll}
\dotxc(t)&=&A\xc(t)+B \sat(v(t)) %\sat(K\xc(t))
&\text{if}\, t\notin \mathcal{M}\\
\xc(t^+)&=&\pi(\xp(t), \xc(t), t)&\text{if}\,t\in\mathcal{M}\\
v(t)&=&K\eta(t)&\, \forall t\in \R_{\geq 0}
\end{array}\right.
\label{eq:Controller}
\end{equation}
where the vector $\xc\in\R^{\np}$ represents the controller state. The matrix $K\in\R^{\nuu\times \np}$ is a control gain to be designed and the function $\pi\colon\R^{2\np}\times\mathcal{M}\rightarrow\R^{\np}$ is defined for all $(\xp, \xc, t)\in\R^{2\np}\times\mathcal{M}$ as
$$
\begin{aligned}
\pi(\xp, \xc, t)\coloneqq&(\pi_1(\xp^{(1)}, \xc^{(1)}, t), \pi_2(\xp^{(2)}, \xc^{(2)}, t),\\
 &\dots, \pi_{\ns}(\xp^{(\ns)}, \xc^{(\ns)}, t))
\end{aligned}
$$
with $\pi_{i}(\xp^{(i)}, \xc^{(i)}, t)\coloneqq
\begin{cases}
\xp^{(i)}&\,\,\mbox{if}\,\,t\in M_i\\
\xc^{(i)}&\,\,\mbox{otherwise}
\end{cases},\qquad \forall i\in\N$. 
The operating principle of the controller in \eqref{eq:Controller} is as follows. The arrival of a new measurement $\xpi$ for some $i\in\N$ triggers an instantaneous jump in the controller state $\eta$ via the function $\pi$. At each jump, the measured vector $\xpi$ is instantaneously stored in $\xci$. If different components of $\xp$ are measured simultaneously, then all the corresponding components of $\xc$ get updated.
Then, in between consecutive measurements, $\eta$ is
continuously updated according to continuous-time dynamics mimicking 
the plant dynamics and its value is continuously used in place of the plant state $\xpi$ in a static state-feedback controller scheme.

The following assumption on the event times is considered.
\begin{assumption}
For each $i\in\N$, the sequence $\{t_k^{(i)}\}^{\infty}_{k=1}\!$ is unbounded and such that there exist two positive real numbers $T^{(i)}_1\leq T_2^{(i)}$  such that  
\begin{equation}
\begin{array}{lr}
\label{eq:timebound}
0\leq t^{(i)}_{1}\leq T_2^{(i)}&\\
T_1^{(i)}\leq t^{(i)}_{k+1}-t^{(i)}_{k}\leq T_2^{(i)}&\qquad\forall k\in\Nats.
\end{array}
\end{equation}
\hfill$\triangle$
\end{assumption}

\begin{remark}
The lower bound in condition \eqref{eq:timebound} prevents the existence of accumulation points in the sequence $\{t_k\}^{\infty}_{k=1}$, and, hence, avoids the existence of Zeno behavior, which is typically undesired in practice. The upper bound in condition \eqref{eq:timebound} guarantees that the time in between samples is bounded above, i.e,. persistent sampling.  
Notice that, the pairs of parameters $(T_1^{(i)}, T_2^{(i)})$ are in general different for different values of $i$. This enables our model to capture asynchronous aperiodic sampling events. The first bound in \eqref{eq:timebound} allows to consider a jump at the initial time, i.e., $t_1=0$.
\hfill$\circ$	
\end{remark}

Our goal is to provide sufficient conditions for exponential stability for the closed-loop system obtained by interconnecting controller \eqref{eq:Controller} with plant \eqref{eq:Plant}.
Then, by defining the error $\tilde{\eta}\coloneqq x_p-\eta$ and by taking as a state $\xcl\coloneqq(\xp, \xctilde)$,
the closed-loop system can be modeled via the following dynamical system with jumps:
\begin{equation}
\left\{\begin{array}{llll}
\hspace{-0.0cm}\dot{\xcl}(t)\hspace{-0.5cm}&=A_{cl}\xcl(t)+\!B_{cl}\dz\left(K_{cl}\xcl(t)\right)&
 t\notin \mathcal{M}\\
\hspace{-0.0cm}\xcl(t^+)\hspace{-0.0cm}&=\hspace{-0.1cm}\begin{bmatrix}
\xp(t)\\
\tilde{\pi}(\xcl(t), t)
\end{bmatrix}&t\in \mathcal{M}
\hspace{-0.3cm}
\end{array}\right.
\label{eq:ClosedLoop}
\end{equation}
where for all $t\in\mathcal{M}, \xcl\in\R^{2\np}$, $\tilde{\pi}(\xcl, t)\coloneqq 
\xp-\pi(\xp, \xp-\xctilde, t)$
and 
$$
\begin{array}{ll}
\!\!\!\!\!\!&A_{cl}\coloneqq\begin{bmatrix}
A+BK&-BK\\
0&A
\end{bmatrix},\, K_{cl}\coloneqq\begin{bmatrix}
K&-K
\end{bmatrix}, B_{cl}\coloneqq\begin{bmatrix}
B\\
0
\end{bmatrix}
\end{array}
$$
and $y\mapsto\dz(y)\coloneqq\sat(y)-y$ is the decentralized deadzone nonlinearity. Notice that 
the definition of $\tilde{\pi}$ ensures that for all $(\xp, \xctilde, t)\in\R^{2\np}\times\mathcal{M}$:
$$
\tilde{\pi}_{i}(\xp^{(i)},\xctilde^{(i)}, t)=
\begin{cases}
0&\,\,\mbox{if}\,\,t\in M_i\\
\xctilde^{(i)}&\,\,\mbox{otherwise}
\end{cases},\qquad \forall i\in\N.
$$
%
%In this paper, we consider the following notion of exponential stability for $\mathcal{H}$.
%\begin{definition}[Exponential stability~\cite{teel2013lyapunov}]
%Given a hybrid system $\mathcal{H}$ with state in $\R^n$ and a closed set $\mathcal{A}\subset\R^{n}$, the set $\mathcal{A}$ is said to be exponentially stable for 
%$\mathcal{H}$ if there exist positive real numbers $\kappa, \lambda$, and $\mu$ such that every maximal solution $\phi$ to $\mathcal{H}$ with $\vert \phi(0,0)\vert_{\mathcal{A}}\leq\mu$ is complete and
%\begin{equation}
%\label{eq:def:les}
%\vert\phi(t, j)\vert_{\mathcal{A}}\leq \kappa e^{-\lambda (t+j)}\vert\phi(0, 0)\vert_{\mathcal{A}}\qquad \forall (t, j)\in\dom\phi
%\end{equation}
%\hfill$\Diamond$
%%If $\mu$ can be selected arbitrarily large, we say that $\mathcal{A}$ is globally exponentially stable for $\mathcal{H}$. 
%\end{definition}
%black
%\begin{definition}[Basin of attraction]
%Let $\mathcal{A}\subset\R^{n}$ be exponentially stable for $\mathcal{H}$. The basin of attraction of $\A$ is the set of points $\mathcal{B}^{\mathcal{H}}_e(\A)\subset\R^{n}$ such that each $\phi\in\mathcal{S}_{\mathcal{H}}(\mathcal{B}_e^\A)$ is complete and satisfies \eqref{eq:def:les}.  \hfill$\Diamond$
%\end{definition}
\begin{remark}
It is worth to mention that controller \eqref{eq:Controller} is not a standard zero-order-holder implementation of a continuous-time controller, but rather a genuinely hybrid controller. 
\end{remark}
\subsection{Hybrid modeling}
\label{sec:Modeling}
Since the closed-loop system \eqref{eq:ClosedLoop}  experiences jumps when a new measurement is available, we rely on the hybrid dynamical model framework in \cite{goebel2012hybrid}. In particular, we provide a single hybrid model capturing the behavior of the closed loop due to each possible evolution generated by any sequence $\{t_k^{(i)}\}^{\infty}_{k=1}\!$, for $i\in\mathcal{N}$,  satisfying \eqref{eq:timebound}. This approach leads to a hybrid system with nonunique solutions allowing one to establish a result for all family of sequences $\{t_k^{(i)}\}^{\infty}_{k=1}\!$, for $i\in\mathcal{N}$, satisfying  \eqref{eq:timebound}.
The proposed modeling approach requires to model the hidden time-driven mechanism triggering the jumps of the controller.
To this end, following \cite{carnevale2007lyapunov}, we add $\ns$ timer variables $\tau_1,\tau_2,\dots, \tau_{\ns}$ to keep track of the duration of flows and to trigger jumps  according to the mechanism in \eqref{eq:ClosedLoop}. 
To accomplish that, we make $\tau_1,\tau_2,\dots, \tau_{\ns}$ decrease as ordinary time $t$ increases and, whenever $\tau_i=0$ for some $i\in\Q$, reset it to any point in $[T_1^{(i)}, T_2^{(i)}]$, so to enforce \eqref{eq:timebound}. In particular, define $\tau\coloneqq(\tau_1,\tau_2,\dots, \tau_{\ns})$, $ \mathcal{T}\coloneqq \bigtimes_{i=1}^{\ns} [0, T_2^{(i)}]$, and for all $i\in\Q, \tau\in\mathcal{T}$
\begin{equation}
\begin{aligned}
\Gt_i(\tau)\coloneqq&\Big\{\chi\in\R^{\ns}\colon \chi_i\in [T_1^{(i)},T_2^{(i)}],\\
&\quad\chi_k=\tau_k \quad\forall k\!\in\!\Q\!\setminus\!\{i\}\Big\}.
\label{eq:Ti}
\end{aligned}
\end{equation}

Then, we consider the following dynamics for $\tau$:
\begin{equation}
\begin{array}{lcll}
\dot{\tau}&=&-\1_{\ns}&\hspace{0.3cm}\tau\in\mathcal{T}\\
\tau^+&\in&\displaystyle{\bigcup_{i\in \{k\in\Q\colon \tau\in \mathcal{D}_k\}}}\!\!\!\!\!\!\!\!\Gt_{i}(\tau)&\hspace{0.3cm}\tau\in\widehat{\mathcal{D}}
\label{eq:DynamicsTimers}
\end{array}
\end{equation}
where $\widehat{\mathcal{D}}\coloneqq\bigcup_{i\in\Q}\D_i$ with, for each $i\in\Q$, $\mathcal{D}_i\coloneqq\{\tau\in\mathcal{T}\colon \tau_i=0\}$. In particular, the dynamics defined in \eqref{eq:DynamicsTimers} enable the timer variable to flow when $\tau\in\T$ and enforce a jump whenever, for some $i=1, 2,\dots, \ns$, $\tau_i=0$. Specifically, at each jump only one component of the vector 
$\tau$ experiences a jump. Indeed, notice that when $\tau\in \D_i\cap \D_j$ for some $i\neq j\in\Q$, only one component of the vector $\tau$ undergoes a change at the jump and multiple consecutive jumps occur. This phenomenon directly follows from having defined an outer semicontinuous jump map\footnote{A set-valued map is outer semicontinuous if its graph is closed; see \cite[Definition 5.4 and Theorem 5.7, pages 152 and 154]{rockafellar2009variational}.}.

We take as state $x\coloneqq(\xcl,\tau)\in\R^{2\np}\times\mathcal{T}$, which allows to consider the state $\xcl$ in \eqref{eq:ClosedLoop} along with the timer variable $\tau$ in \eqref{eq:DynamicsTimers} and define the flow map and the flow set as follows: 
\begin{equation}
f(x)\coloneqq\begin{bmatrix}
A_{cl}\xcl+B_{cl}\dz(K_{cl}\xcl)\\
-\1_{\ns}
\end{bmatrix}, \quad\forall x\in
\mathcal{C}\coloneqq\R^{2\np}\times\mathcal{T}.
\label{eq:F}
\end{equation}
To define the jump map, as a first step, define for all $i\in\Q, \xctilde\in\R^{\np}$, $
\gxc_i(\tilde{\eta})\coloneqq\left(\displaystyle\bigoplus_{j=1}^{\ns} H^{(i)}_j\right) \tilde{\eta}$,
where for each $i, j\in\Q$,
$
H^{(i)}_j\coloneqq\begin{cases}
\0_{n_j\times n_{j}}&\,\,\mbox{if}\,\,i=j\\
\Id_{n_j}&\,\,\mbox{otherwise}
\end{cases}
$. Finally, from (\ref{eq:Ti}) and  \eqref{eq:DynamicsTimers}, one can define the jump map and the jump set as follows:
\begin{equation}
G(x)\coloneqq \begin{bmatrix}
x_p\\
 \mathcal{J}(\tilde{\eta},\tau)
 \end{bmatrix},\quad\forall x\in\D\coloneqq \R^{2\np}\times\widehat{\mathcal{D}} 
\label{eq:G}
\end{equation}
with
$
\displaystyle\mathcal{J}(\tilde{\eta},\tau)\coloneqq \bigcup_{i\in \{k\in\Q\colon \tau\in \mathcal{D}_k\}}\left(\{\gxc_i(\tilde{\eta})\}\times \widehat{\mathcal{G}}_{i}(\tau)\right)$.
The jump map in (\ref{eq:G}) enables to capture the dynamics of the closed-loop system at the sampling times. Indeed, for all $x\in\D$, one has that $\tilde{\eta}^+_{i}=0$ if $\tau_i=0$, and $\tilde{\eta}^+_{i}=\tilde{\eta}_{i}$ otherwise.

Therefore, with the definitions in (\ref{eq:F}) and  (\ref{eq:G}), the closed-loop system reads:
\begin{equation}
\label{eq:HybridCL}
%\mathcal{H}_{cl}
\left\{
\begin{array}{ccll}
\dot{x}&=&f(x)&\quad x\in \C\\
x^+&\in&G(x)&\quad x\in \D.
\end{array}\right.
\end{equation}
%%%%%%%%
The following lemma shows that \eqref{eq:HybridCL} is well-posed\footnote{A hybrid system, like (\ref{eq:HybridCL}), is said to be well-posed if its set of solutions has good structural properties; see \cite[Definition 6.27]{goebel2012hybrid}. In simple words, well-posed hybrid systems are such that, in particular, every sequence of bounded solutions with converging initial conditions converge to a solution in the graphical sense.}  thereby ensuring robustness with respect to small perturbations; see \cite{goebel2012hybrid} for more details. 
\begin{lemma}
\label{lemm:wellposed}
The sets $\C$ and $\D$ are closed, $f$ is continuous, and $G$ is locally bounded and outer semicontinuous on $\D$. Consequently, the closed-loop system (\ref{eq:HybridCL}) is well-posed.
\end{lemma}
\begin{pf}
First observe that the flow set $\C$ and the jump set $\D$ are closed since $\T$ is closed. Moreover, the flow map $f$ is continuous since $\dz$ is continuous. Hence, according to \cite[Theorem 6.8]{goebel2012hybrid} well-posedness of $\Hcl$ follows if the set-valued map $G$ in \eqref{eq:G} is outer semicontinuous and locally bounded on $\D$. Now we show that these properties hold. 
 
First we show that $G$ is outer semicontinuous. To this end, thanks to  \cite[Lemma 5.10]{goebel2012hybrid}, it suffices to show that $\gph G$ is closed.
Consider the following rewriting of  $G$ 
$$
G(x)=\bigcup_{i\in \{k\in\Q\colon \tau\in \mathcal{D}_k\}}\widetilde{G}_i(x)
 $$
where, for each $i\in\N$,
$$
 \widetilde{G}_i(x)\coloneqq\{x_p\}\times \{\gxc_i(\tilde{\eta})\}\times \Gt_{i}(\tau)\qquad \forall (\xp, \xctilde, \tau)\in\D
$$
In particular, notice that for each $i\in\N$, $\gph\Gt_i$ is closed;  this property follows immediately by noticing that for each $i\in\N$
$$
\gph \Gt_i= \left\{(z, r)\in\R^{2\ns}\colon (z_i, r_i)\in\R\times[T^{(i)}_1, T^{(i)}_2] \right\}
$$ 
The latter implies that for each $i\in\N$, $\gph\widetilde{G}_i$ is closed. Indeed, by denoting $\id\colon\R^{n}\rightarrow\R^{n}$ the identity function, one has  
$$
\gph\widetilde{G}_i=\gph\id\times\gph \gxc_i\times\gph \Gt_i
$$
which is closed, $\gxc_i$ being a continuous function and $\gph \Gt_i$ being closed for each $i\in\N$. Hence, $\gph G$, being the finite union of closed sets, is closed implying that $G$ is outer semicontinuous on $\D$.

To show that $G$ is locally bounded on $\D$ first notice that for each $x\in\D$
$$
G(x)\subset \bigcup_{i=1}^{\ns}\left(
 \{x_p\}\times\{ \gxc_i(\tilde{\eta})\}\times \Gt_i(\tau)\right)=\colon\Gamma(x)
$$
Let $S\subset\D$ be bounded, then
$$
\begin{aligned}
&G(S)\subset\Gamma(S)\subset\\
&\bigcup_{i=1}^{\ns}\left(
\Pi_{\xp}(S)\times \gxc_i(\Pi_{\tilde{\eta}}(S))\times \Gt_i(\Pi_{\tau}(S))\right)
\end{aligned}
 $$
where
$$\scalebox{0.95}{$
\begin{aligned}
&\Pi_{\xp}(S)\coloneqq\{\xp\in\R^{n}\colon \exists (\xctilde, \tau)\in \R^{n+\ns}\,\text{s.t.}\,\, (\xp, \xctilde, \tau)\!\in S\}\\
&\Pi_{\xctilde}(S)\coloneqq\{\xctilde\in\R^{n}\colon \exists (\xp, \tau)\in \R^{n+\ns}\,\text{s.t.}\,\,(\xp, \xctilde, \tau)\in S\}\\
&\Pi_{\tau}(S)\coloneqq\{\tau\in\R^{\ns}\colon \exists (\xp, \xctilde)\in \R^{2n}\,\text{s.t.}\,\, (\xp, \xctilde, \tau)\in S\}
\end{aligned}$}
$$
Thus, since $\Gt_i$ and $\gxc_i$ are locally bounded on $\D$ for each $i\in \N$, it follows that $G$ is locally bounded on $\D$. This concludes the proof. \QEDA
\end{pf}

Existence of solutions to system (\ref{eq:HybridCL}) follows directly from the definition of the system, by checking that for every initial condition $\xi\in\mathcal{C}\cup \mathcal{D}$ there exists a nontrivial solution and every maximal solution to (\ref{eq:HybridCL}) from $\xi$ is complete.
Moreover,  the definition of (\ref{eq:HybridCL}), similarly as in \cite{ferrante2015ACC}, ensures that for each solution, at most $\ns$ jumps can occur consecutively without flowing.  Such a property, along with \eqref{eq:timebound}, ensures that for every maximal solution $\phi$ to  (\ref{eq:HybridCL}) and each $(t, j)\in\dom \phi$ such that  $(t, s)\in\dom\phi$ for some $s\in\{j+1,j+2,\dots, j+\ns\}$, one has $\left([t,t+\min\{T_1^{(1)},T_1^{(2)},\dots, T_1^{(\ns)}\}]\times \{s\}\right)\subset\dom\phi$. Essentially, the domain of the solutions to (\ref{eq:HybridCL})  manifests an average dwell-time property, with dwell time $\tau_D=\min\{T_1^{(1)},T_1^{(2)},\dots, T_1^{(\ns)}\}$ and offset $N_0=\ns$,  which implies for each $(t,j)\in\dom\phi$ 
\begin{equation}
\label{eq:TimeBound}
\tau_D j\leq t+\ns\tau_D
\end{equation}
see, e.g., \cite[Example 2.15]{goebel2012hybrid}. Such a property prevents from the existence of Zeno solutions.

The goal of this paper is to characterize exponential stability of the origin for the $\xcl$ substate of the closed loop (\ref{eq:HybridCL}), uniformly in $\tau$. The fact that $\tau$ evolves in the compact set $\mathcal{T}$ simplifies this task and allows stating it as a suitable stability property for the compact set
\begin{equation}
\label{eq:P2:Chap3:A}
\mathcal{A}\coloneqq\{0\}\times \T\subset\R^{2\np}\times\R_{\geq 0}^{\ns}
\end{equation}
To this end, we adopt the notion of exponential stability of closed sets for hybrid systems as introduced in \cite{teel2013lyapunov} and recalled next.
\begin{definition}[Exponential stability~\cite{teel2013lyapunov}]
The set $\mathcal{A}$ is said to be exponentially stable for 
(\ref{eq:HybridCL}) if there exist positive real numbers $\kappa, \lambda$, and $\mu$ such that every maximal solution $\phi$ to (\ref{eq:HybridCL}) with $\vert \phi(0,0)\vert_{\mathcal{A}}\leq\mu$ is complete and it satisfies
\begin{equation}
\label{eq:def:les}
\vert\phi(t, j)\vert_{\mathcal{A}}\leq \kappa e^{-\lambda (t+j)}\vert\phi(0, 0)\vert_{\mathcal{A}}\qquad \forall (t, j)\in\dom\phi.
\end{equation}
The set of points, denoted by $\mathcal{B}(\A)$, such that each maximal solution to (\ref{eq:HybridCL}) from $\mathcal{B}(\A)$ is complete and converges to $\A$ is denoted as the basin of  
attraction of 
$\A$. 
\hfill$\Diamond$
\end{definition}
It is important to note that due to the presence of input saturation, the implicit objective is to characterize the basin of attraction of the closed set for which the exponential stability is guaranteed. Indeed, the basin of attraction can be $\R^{2\np}\times \T$ only if the open-loop system, namely matrix $A$ has some particular stability property; see \cite{tarbouriech2011stability} for these questions. 
The problem we solve is formalized as follows.
\begin{problem}[Controller Design] 
\label{prob:Problem1}
Given matrices $A$ and $B$ of appropriate dimensions, and positive real numbers $T_1^{(1)}, T_2^{(1)}, \dots, T_1^{(\ns)}, T_2^{(\ns)}$, design $K\in\R^{\nuu\times \np}$ such that the set 
$\mathcal{A}$ defined in \eqref{eq:P2:Chap3:A} is exponentially stable for for the hybrid system \eqref{eq:HybridCL} and provide an inner approximation of the basin of attraction of $\A$, i.e., a subset of the basin of attraction.  \hfill$\diamond$ 
\end{problem}

\section{Lyapunov-Based Stability Analysis}
\label{sec:Lyap}

To address Problem \ref{prob:Problem1}, we use the generalized sector condition for the deadzone nonlinearity presented 
in \cite{tar:pri:gom/ieee06}, which is recalled next in a convenient form.

\begin{lemma}
\label{lemm:sector}
Let $X\in\Dy^{\nuu}_+$, $L\in\R^{\nuu\times 2\np}$, and $z\in\R^{2\np}$ be such that for all $i\in\{1,2,\dots,\nuu\}$, $\vert L_{(i)}z\vert\leq \bar{u}_{i}$, where $\bar{u}_1, \bar{u}_2,\dots, \bar{u}_\nuu$ are defined in Section~\ref{sec:ProbStat}.  
Then, for each $u\in\R^{\nuu}$:
\begin{equation}
\label{eq:sec:loc}
\dz(u)\tr X(\dz(u)+u+Lz)\leq 0.
\end{equation}
\QEDB
\end{lemma}
To exploit the above result in our setting, it is convenient to state the following preliminary result.
\begin{lemma}
\label{lemm:inclusion}
Let $\bar{\mu}, \sigma_1, \sigma_2,\dots, \sigma_{\ns}\in\R_{>0}$, $R_1\in\S^{n_{1}}_+, R_2\in\S^{n_2}_+,\dots,$
$R_{\ns}\in\S^{n_\ns}_+$, $W\in\S^{\np}_+$, $Z, J\in\R^{\nuu\times \np}$. Define:
\begin{subequations}
\begin{equation}
\label{eq:E_rho}
\begin{aligned}
&\mathcal{L}\coloneqq \left\{\xcl\in\R^{2 \np} \colon \vert L_{(i)}\xcl\vert\leq \bar{u}_i, i\in\{1,2,\dots,\nuu\}\right\}\\
&\mathcal{Q}_{\bar{\mu}}\coloneqq\left\{x=(\xcl, \tau)\in\C \colon \xcl\tr\widehat{P}(\tau)\xcl\leq \bar{\mu}\right \}
\end{aligned}
\end{equation}
where for all $\tau\in\T$
\begin{equation}
\begin{aligned}
&L\coloneqq\begin{bmatrix}
ZW^{-1}&J
\end{bmatrix},
&\widehat{P}(\tau)\coloneqq W^{-1}\oplus R(\tau),\\
&R(\tau)\coloneqq\bigoplus_{i=1}^{\ns}R_{i}e^{\sigma_i \tau_i}.
\end{aligned}
\label{eq:lemma:Phat}
\end{equation}
\end{subequations}
Assume that
\begin{equation}
\label{eq:lemm:PolyInclusion}
\begin{bmatrix}
W&\0&Z\tr_{(i)}\\
\star&\displaystyle\bigoplus_{j=1}^{\ns}R_j&J\tr_{(i)}\\ 
\star&\star&\frac{\bar{u}^2_i}{\bar{\mu}}
\end{bmatrix}\succeq 0\qquad \forall i\in\{1,2,\dots,\nuu\}.
\end{equation}
Then, the following inclusion holds:
\begin{equation}
\label{eq:lemma:inclusion}
\mathcal{Q}_{\bar{\mu}}\subset \mathcal{L}\times\T.
\end{equation}
\end{lemma}
\medskip
\begin{pf}
Define the following sets
\begin{equation}
\begin{aligned}
&\mathcal{R}\coloneqq\left\{\xcl\in\R^{2\np} \colon \xcl\tr L_{(i)}\tr L_{(i)}\xcl\leq \bar{u}^2_i, i\in\{1,2,\dots, \nuu\}\right\}\\
&\mathcal{E}^{0}_{\bar{\mu}}\coloneqq\left\{\xcl\in\R^{2\np} \colon \xcl\tr \widehat{P}(0)\xcl\leq \bar{\mu}\right\}
\end{aligned}
\label{eq:Erho0}
\end{equation}
and observe that $\mathcal{R}\subset\mathcal{L}$. Let $\mathbb{J}\coloneqq W^{-1}\bigoplus\Id\bigoplus\Id$, then by pre-and-post multiplying the left-hand side of \eqref{eq:lemm:PolyInclusion} by $\mathbb{J}$ one gets
\begin{equation}
\label{eq:lemma:congruence}
\begin{bmatrix}
W^{-1}&\0&W^{-1}Z\tr_{(i)}\\
\star&\displaystyle\bigoplus_{j=1}^{\ns}R_j&J\tr_{(i)}\\ 
\star&\star&\frac{\bar{u}^2_i}{\bar{\mu}}
\end{bmatrix}\succeq 0\qquad \forall i\in\{1,2,\dots,\nuu\}.
\end{equation}  
By Schur complement, \eqref{eq:lemma:congruence} yields, for all $i\in\{1,2,\dots,\nuu\}$
$
\frac{1}{\bar{\mu}}\widehat{P}(0)-\frac{1}{\bar{u}^2_i}L_{(i)}\tr L_{(i)}\succeq 0
$,
that is $\mathcal{E}^{0}_{\bar{\mu}}\subset\mathcal{R}$, which in turn implies 
\begin{equation}
\mathcal{E}^{0}_{\bar{\mu}}\subset\mathcal{L}.
\label{lemma:Poly:EsubG}
\end{equation}
Now observe that for each $\tau\in\T$, one has $\widehat{P}(\tau)\succeq\widehat{P}(0)$. Therefore $\mathcal{Q}_{\bar{\mu}}\subset \mathcal{E}_{\bar{\mu}}^{0}\times\T$
which, thanks to \eqref{lemma:Poly:EsubG}, implies that $\mathcal{Q}_{\bar{\mu}}\subset \mathcal{L}\times\T$. 
This concludes the proof.\QEDA
\end{pf}
\begin{remark}
Lemma~\ref{lemm:inclusion} enables to exploit the sector condition in Lemma~\ref{lemm:sector}, which is typically coupled with quadratic Lyapunov functions, 
with ``clock-dependent'' Lyapunov functions. This key aspect of the paper clearly emerges in Theorem~\ref{theo:ExpStab}.\hfill$\circ$
\end{remark}

Now we are in a position to state the main result, which provides sufficient conditions for exponential stability of the closed-loop system in \eqref{eq:HybridCL} together with
an explicit estimate of the basin of attraction of $\A$.
\begin{theorem}[Exponential stability]
\label{theo:ExpStab}
Let $K\in\R^{\nuu\times \np}$ be given. Suppose that there exist $W\in\S^{\np}_+$, $R_1\in\S^{n_1}_+, R_2\in\S^{n_2}_+, \dots,R_{\ns}\in\S^{n_{\ns}}_+$, $S\in\Dy^{\nuu}_+$, $\sigma_1, \sigma_2,\dots, \sigma_{\ns}, \bar{\mu}\in\R_{>0}$, $Z, J\in\R^{\nuu\times \np}$ such that \eqref{eq:lemm:PolyInclusion} holds and that for all $\tau\in\T$ 
\begin{equation}
\label{eq:thm:LMI_LES}
\scalebox{0.85}{$
\underbrace{\begin{bmatrix}
\He((A+BK)W)&-BK&BS-WK\tr- Z\tr\\
\star&\He(R(\tau)A)-\Sigma R(\tau)&K\tr-J\tr\\
\star&\star&-2S
\end{bmatrix}}_{\mathfrak{M}(\tau)}\!\prec\!0$},
\end{equation}
where, for all $\tau\in\R^{\ns}$, $R(\tau)\coloneqq\displaystyle\bigoplus_{j=1}^{\ns}e^{\sigma_j\tau_j}R_j$ and $\Sigma\coloneqq\displaystyle\bigoplus_{j=1}^{\ns}\sigma_j\Id_{n_j}$. Then, the set $\A$ defined in \eqref{eq:P2:Chap3:A} is exponentially stable for (\ref{eq:HybridCL})  and the set $\mathcal{Q}_{\bar{\mu}}$ defined in \eqref{eq:E_rho}
is included in the basin of attraction of $\A$. In particular, there exist positive numbers $\lambda$ and $\kappa$ such that for any maximal solution $\phi$ to (\ref{eq:HybridCL}), $\phi(0, 0)\in \mathcal{Q}_{\bar{\mu}}$ implies
$$
\vert \phi(t, j)\vert_{\A}\leq \kappa e^{-\lambda(t+j)}\vert \phi(0, 0)\vert_{\A}\quad\forall (t, j)\dom\phi.
$$
\end{theorem}
\begin{pf}
The proof of the result hinges upon Theorem~\ref{theo:LES} by showing that the hypotheses of the result imply all the conditions in Theorem~\ref{theo:LES}.

Inspired by \cite{ferrante2018mathcal,forni2014event}, for each $x\in\R^{\nx}$ define
\begin{equation}
\label{eq:V}
V(x)=\xp\tr W^{-1} \xp+\xctilde\tr R(\tau) \xctilde=:\xcl\tr \widehat{P}(\tau)\xcl
\end{equation}
where for all $\tau\in\R^{\ns}$, $R(\tau)\coloneqq R_{i}e^{\sigma_i \tau_i}$. 
Notice that $V$ is continuously differentiable on $\R^{\nx}$ and that for all $x\in\C$, $c_1\vert x\vert_\A^2\leq V(x)\leq c_2\vert x\vert_\A^2$, with $c_2\coloneqq\displaystyle\max_{\tau\in\T}\lambda_{\max}(\widehat{P}(\tau))$ and $c_1\coloneqq\displaystyle\min_{\tau\in\T}\lambda_{\min}(\widehat{P}(\tau))$, which are both strictly positive due to $P, R_1, \dots, R_{\ns}$ being positive definite\footnote{By the definition of the system  (\ref{eq:HybridCL}) and of the set $\mathcal{A}$, for every $x\in \C \cup \D \cup G(\D)$, $\vert x \vert_\mathcal{A}=\vert (\xp, \xctilde) \vert$.} . In addition, from Lemma~\ref{lemm:V}, one has that for all $x\in\D, g\in G(x)$, $V(g)-V(x)\leq 0$. Let $\mathcal{L}$ be defined as in \eqref{eq:E_rho}. We now show that the satisfaction of \eqref{eq:thm:LMI_LES} implies a suitable decreasing property for $V$ during flows.
Pick any $\xcl\in\R^{2\np}$ and $\tau\in\R^{\ns}$, then
$
\left\langle\frac{\partial}{\partial\tau}\xctilde\tr R(\tau)\xctilde , -\1_{\ns}\right\rangle=-\xctilde\tr\Sigma R(\tau)\xctilde.
$
Thus, for all $x\in\C$, one has
%Thus, one has 
\begin{equation}
\begin{array}{ll}
\langle\nabla V(x), f(x)\rangle=&\xp\tr\He(W^{-1}(A+BK))\xp\\
&+2\xp\tr W^{-1} B\dz(K_{cl}\xcl)\\
&+\xctilde\tr\He(R(\tau)A)\xctilde\\
&-\xctilde\tr\Sigma R(\tau)\xctilde-2\xp\tr W^{-1}BK\xctilde.
\end{array}
\end{equation}
The proof is now completed by using an S-procedure argument; see, e.g., \cite{Boyd}.
Specifically, pick any $x\in\mathcal{L}\times\T$. Then, from Lemma~\ref{lemm:sector}, the satisfaction of \eqref{eq:lemm:PolyInclusion} implies that
\begin{equation}
\langle\nabla V(x), f(x)\rangle\leq \langle\nabla V(x), f(x)\rangle-2\Omega(\xcl)
\label{eq:VdotBoundRob_1}
\end{equation}
where for all $\xcl\in\R^{2\np}$
\begin{equation}
\Omega(\xcl)\coloneqq\dz(K_{cl}\xcl)\tr S^{-1}(\dz(K_{cl}\xcl)+K_{cl}\xcl+L\xcl)
\label{eq:Omega}
\end{equation}
and $L$ is defined in \eqref{eq:lemma:Phat}. Let for all $\xcl\in\R^{2\np}$, $\varsigma(\xcl)\coloneqq(\xcl, \dz(K_{cl}\xcl))$. Straightforward calculations show that for all $x\in\C$, the right-hand
side of \eqref{eq:VdotBoundRob_1} can be rewritten as
$
\varsigma(\xcl)\tr 
\underbrace{\left[\begin{smallmatrix}
W^{-1}&0&0\\
\star&\Id&0\\ 
\star&\star&S^{-1}
\end{smallmatrix}\right]
\mathfrak{M}(\tau)\left[\begin{smallmatrix}
W^{-1}&0&0\\
\star&\Id&0\\ 
\star&\star&S^{-1}
\end{smallmatrix}\right]}_{\widehat{\mathfrak{M}}(\tau)}\varsigma(\xcl)
$
with $\mathfrak{M}(\tau)$ defined in \eqref{eq:thm:LMI_LES}. Namely, since \eqref{eq:VdotBoundRob_1} holds for any $x\in\mathcal{L}\times\T$, one has
\begin{equation}
\langle\nabla V(x), f(x)\rangle\leq \varsigma(\xcl)\tr \widehat{\mathfrak{M}}(\tau)\varsigma(\xcl)\quad\forall x\in\mathcal{L}\times\T.
\label{eq:VdotBoundRob_2}
\end{equation}
Pick  $c_3\coloneqq -\displaystyle{\max_{\tau\in\T}}\lambda_{\max}(\widehat{\mathfrak{M}}(\tau))$,
which is well defined due to the entries of $\widehat{\mathfrak{M}}$ being continuous functions of $\tau$ and strictly positive thanks to the satisfaction of \eqref{eq:thm:LMI_LES}. Then, using \eqref{eq:VdotBoundRob_2} one gets:
\begin{equation}
\langle\nabla V(x), f(x)\rangle\leq-c_3 \xcl\tr \xcl\leq-\frac{c_3}{c_2}V(x)\quad \forall x\in\mathcal{L}\times\T,
\label{eq:Vdot_Bound_M}
\end{equation}
where the last bound follows from the sandwich inequality shown earlier in the proof and the definition of $\A$. 
Thus, $V$ satisfies all the conditions in Theorem~\ref{theo:LES} with $\mathcal{U}=\mathcal{L}$. Therefore, the set $\mathcal{A}$ defined in \eqref{eq:P2:Chap3:A} is exponentially stable for (\ref{eq:HybridCL})  and each sublevel set of $V$ included in $\mathcal{L}\times\T$ is contained in the basin of attraction of $\A$. 
To conclude the proof, observe that $\mathcal{Q}_{\bar{\mu}}=L_{\left.V\right\vert_{\dom V\cap C}}(\bar{\mu})$ and that thanks to Lemma~\ref{lemm:inclusion}, \eqref{eq:lemm:PolyInclusion} implies that $\mathcal{Q}_{\bar{\mu}}\subset\mathcal{L}\times\T$. This concludes the proof. \QEDA
\end{pf}
\section{Control Design}
\label{subsec:ControlDesign}
This section devises a computationally tractable procedure for designing the controller gain $K$.

\subsection{Computationally tractable sufficient conditions}
\label{sec:StabAnalysis}
%%%%%Recasting into a finite number of LMIs
Theorem~\ref{theo:ExpStab} turns the stability analysis problem of the closed-loop system into the feasibility problem of a collection of matrix inequalities. Nonetheless, the search for feasible solutions to \eqref{eq:thm:LMI_LES} is quite challenging since such conditions need to be checked over an uncountable set, \ie, $\T$. To overcome this problem, next we show that \eqref{eq:thm:LMI_LES} can be turned without any additional conservatism into a finite collection of matrix inequalities. To this end, first consider the following preliminary result.
\begin{lemma}
\label{lemm:Tech}
Let $\sigma_1,\sigma_2,\dots,\sigma_{\ns}$ be given real positive numbers and for all $\tau\in\T$ define:
	\begin{equation}
	\label{eq:Theta}
	\Theta(\tau)\coloneqq\bigoplus_{i=1}^{\ns} e^{\sigma_i\tau_i}\Id_{n_i}.
	\end{equation}
Then, $\range\Theta$ is a polytope and in particular
	\begin{equation}
	\label{eq:Rrange}
	\range\Theta=\Co\left\{\bigoplus_{i=1}^{\ns} \psi_i \Id_{n_i}\colon \psi_i\in\{1, e^{\sigma_i T_2^{(i)}}\}.\right\}
	\end{equation}
%	\QEDB
\end{lemma}	
\begin{pf}
As a first step, we rewrite $\range\Theta$ in an equivalent fashion. In particular, consider the following linear application: 
$$
\begin{aligned}
g\colon &\R^{\ns}\rightarrow\R^{n\times n}\\
&v\mapsto \bigoplus_{i=1}^{\ns} v_i\Id_{n_{p_{i}}}
\end{aligned}
$$
and let  $\mathcal{W}\coloneqq\bigtimes_{i=1}^{\ns}\Co\{1, e^{\sigma_i T_2^{(i)}}\}$.
Then, it is easy to see that 
\begin{equation}
\label{eq:RangeCalW}
\range\Theta=g(\mathcal{W})
\end{equation}
As a second step, observe that\footnote{We used the following property: Let $S_i\subset\R^{n_i}$ for $i=1,2,\dots, s$. Then, $\displaystyle\Co\bigtimes_{i=1}^sS_i=\bigtimes_{i=1}^s\Co S_i$; see 
\cite{bertsekas2009convex}.}
$$
\mathcal{W}=\Co\bigtimes_{i=1}^{\ns}\{1, e^{\sigma_i T_2^{(i)}}\}
$$
The above relationship, along with \eqref{eq:RangeCalW}, gives
\begin{equation}
\label{eq:RangeTh}
\range\Theta=g\left(\Co\bigtimes_{i=1}^{\ns}\{1, e^{\sigma_i T_2^{(i)}}\}
\right)
\end{equation}
Thanks to Lemma~\ref{lemm:ConvLin} in the Appendix, since the application $g$ is linear, it follows that for all $U\subset\R^{\ns}$ one has 
$g(\Co U)=\Co g(U)$. Hence, form \eqref{eq:RangeTh} it follows that
$$
\range\Theta=\Co g\left(\bigtimes_{i=1}^{\ns}\{1, e^{\sigma_i T_2^{(i)}}\}\right)
$$
which corresponds to \eqref{eq:Rrange}. This concludes the proof. \QEDA
\end{pf}
Based on Lemma \ref{lemm:Tech}, the following proposition shows that \eqref{eq:thm:LMI_LES} can be turned into a finite set of matrix inequalities.
\begin{proposition}
Let $W\in\S^{\np}_+$, $K\in\R^{\np\times \nuu}$, $R_1\in\S^{n_1}_+, R_2\in\S^{n_2}_+, \dots, R_{\ns}\in\S^{n_{\ns}}_+$, $S\in\Dy^{\nuu}_+$, $\sigma_1$, $\sigma_2,\dots, \sigma_{\ns}\in\R_{>0}$, $Z, J\in\R^{\nuu\times \np}$. Let for all $\tau\in\R^{\ns}$,
$R(\tau)\coloneqq\bigoplus_{j=1}^{\ns}e^{\sigma_j\tau_j}R_j$, $\Sigma\coloneqq\bigoplus_{j=1}^{\ns}\sigma_j\Id_{n_{p_j}}$, and $\widehat{R}\coloneqq\bigoplus_{i=1}^{\ns} R_i$. Then, \eqref{eq:thm:LMI_LES} holds if and only if for all $\Psi\in\mathcal{Z}$:
\begin{subequations}
\begin{equation}
\begin{aligned}
&\hspace{0cm}\mathfrak{N}(\Psi)\hspace{-0.1cm}\coloneqq\hspace{-0.1cm}
\left[\begin{smallmatrix}
\He((A+BK)W)&-BK&BS-WK\tr- Z\tr\\
\star&\He(\widehat{R}\Psi A)-\Sigma \widehat{R}\Psi&K\tr-J\tr\\
\star&\star&-2S
\end{smallmatrix}\right]
\!\!\prec 0\\
\label{eq:prop:M_ConvexHull_LES}
\end{aligned}
\end{equation}
where:
\begin{equation}
\label{eq:calZ}
\mathcal{Z}\coloneqq\left\{\bigoplus_{i=1}^{\ns}\psi_i\Id_{n_i}\colon \psi_i\in\{1,e^{\sigma_i T_2^{(i)}}\}\right\}.
\end{equation}
\end{subequations}
\label{prop:FiniteNMI}
\end{proposition}
\begin{pf}
Since for all $\tau\in\T$, $R(\tau)=\widehat{R}\Theta(\tau)$ where $\Theta(\tau)$ is defined in \eqref{eq:Theta},  from \eqref{eq:Rrange} it  
follows that \eqref{eq:thm:LMI_LES} holds if and only if $-\mathfrak{N}(\range\Theta)\subset\S_{>0}$, 
where $\S_{>0}$ is the set of positive definite matrices of adequate dimensions. Thanks to Lemma~\ref{lemm:Tech}, the latter is equivalent to require that:
\begin{equation}
\label{eq:prop:proof:CalN}
-\mathfrak{N}\left(\Co\left\{\bigoplus_{i=1}^{\ns} \psi_i \Id_{n_i}\colon \psi_i\in\{1, e^{\sigma_i T_2^{(i)}}\}\right\}\right)\subset\S_{>0}.
\end{equation}
To conclude, it suffices to notice that since the application $\Psi\mapsto\mathfrak{N}(\Psi)$ is linear, from  Lemma~\ref{lemm:ConvLin} it follows that
 $$
 \begin{aligned}
&\mathfrak{N}\left(\Co\left\{\bigoplus_{i=1}^{\ns} \psi_i \Id_{n_i}\colon \psi_i\in\{1, e^{\sigma_i T_2^{(i)}}\}\right\}\right)=\\
&\qquad\qquad\qquad\,\Co\mathfrak{N}\left(\left\{\bigoplus_{i=1}^{\ns} \psi_i \Id_{n_i}\colon \psi_i\in\{1, e^{\sigma_i T_2^{(i)}}\}\right\}\right). \end{aligned}
 $$
Thus, $\S_{>0}$ being convex, the latter implies that \eqref{eq:prop:proof:CalN} (or equivalently \eqref{eq:thm:LMI_LES}) holds if and only \eqref{eq:prop:M_ConvexHull_LES} holds. This concludes the proof. \QEDA
\end{pf}
The result given next enables to solve Problem~\ref{prob:Problem1} via the solution to some linear matrix inequalities (\emph{LMIs}) coupled to a line search on a few scalar parameters.
\begin{proposition}
\label{prop:ExpStabControl}
Let $W\in\S^{\np}_+$, $Y\in\R^{\np\times \nuu}$, $R_1\in\S^{n_1}_+, R_2\in\S^{n_2}_+, \dots,R_{\ns}\in\S^{n_\ns}_+$, $S\in\Dy^{\nuu}_+$, $\sigma_1, \sigma_2,\dots$, $ \sigma_{\ns}$, $\bar{\mu}, \alpha\in\R_{>0}$, and $Z\in\R^{\nuu\times \np}$. Define $\widehat{R}\coloneqq\displaystyle\bigoplus_{i=1}^{\ns} R_i$
and $\Sigma\displaystyle\coloneqq\bigoplus_{j=1}^{\ns} \sigma_j\Id_{n_{p_j}}$.
If 
\begin{subequations} 
\begin{equation}
\label{eq:prop:PolyInclusion_Design}
\begin{bmatrix}
W&\0&Z\tr_{(i)}\\
\star&\widehat{R}&0\\ 
\star&\star&\frac{\bar{u}^2_i}{\bar{\mu}}
\end{bmatrix}\succeq 0\qquad \forall i\in\{1,2,\dots,\nuu\}
\end{equation}
and for all $\Psi\in\mathcal{Z}$, with $\mathcal{Z}$ defined as in \eqref{eq:calZ}
\begin{equation}
\label{eq:prop:LMI_LES_des}
\scalebox{0.8}{$
\begin{bmatrix}
\He(AW+BY)&-BY&BS-Y\tr-Z\tr&0\\
\star&-2\alpha W&Y\tr&\alpha\Id\\
\star&\star&-2S&0\\
\star&\star&\star&\He(\widehat{R}\Psi A-\Sigma\widehat{R}\Psi)
\end{bmatrix}\prec 0$.}
\end{equation} 
\end{subequations} 
Then, $K=YW^{-1}$ solves Problem~\ref{prob:Problem1}. In particular, 
the set $\A$ defined in \eqref{eq:P2:Chap3:A} is exponentially stable for (\ref{eq:HybridCL})  and the set $\mathcal{Q}_{\bar{\mu}}$ defined in \eqref{eq:E_rho}
is included in the basin of attraction of $\A$.
\end{proposition}

\begin{pf}
As a first step, notice that \eqref{eq:prop:PolyInclusion_Design} corresponds to
\eqref{eq:lemm:PolyInclusion} with $J=0$. Moreover, following the same arguments as in the proof of Proposition~\ref{prop:FiniteNMI} and using $Y=KW$, it follows that
 \eqref{eq:prop:LMI_LES_des} implies that for all $\tau\in\T$:
\begin{equation}
\label{eq:prop:LMI_LES_des_int}
\scalebox{0.85}{$
\begin{bmatrix}
\He((A+BK)W)&-BKW&BS-WK\tr-Z\tr&0\\
\star&-2\alpha W&WK\tr&\alpha\Id\\
\star&\star&-2S&0\\
\star&\star&\star&-Q(\tau)
\end{bmatrix}\prec 0.$}
\end{equation}
where for all $\tau\in\T$, $Q(\tau)\coloneqq -(\He(R(\tau)A-\Sigma R(\tau)))$. Via Schur complement, the latter gives, for all $\tau\in\T$:
\begin{equation}
\scalebox{0.85}{$
\begin{bmatrix}
\He((A+BK)W)&-BKW&BS-WK\tr-Z\tr\\
\star&-2\alpha W+\alpha^2Q^{-1}(\tau)&WK\tr\\
\star&\star&-2S
\end{bmatrix}\prec 0.$}
\label{eq:PartialDesignMI}
\end{equation}
Now observe that from \eqref{eq:prop:LMI_LES_des_int}, it follows that for all $\tau\in\T$, $Q(\tau)\succ 0$. Thus, by recalling the following elementary inequality
\begin{equation}
\label{eq:BoundW}
(\alpha Q(\tau)^{-1}-W) Q(\tau) (\alpha Q(\tau)^{-1}-W)\succeq 0 \quad \forall\tau\in\T
\end{equation}
one has that for all $\tau\in\T$, $-WQ(\tau)W\preceq\alpha^2 Q^{-1}(\tau)-2\alpha W$. Therefore, using \eqref{eq:PartialDesignMI} one has, for all $\tau\in\T$
$$
\scalebox{0.85}{$
\begin{bmatrix}
\He((A+BK)W)&-BKW&BS-WK\tr-Z\tr\\
\star&-WQ(\tau)W&WK\tr\\
\star&\star&-2S
\end{bmatrix}\prec 0.$}
$$
Finally, by pre-and-post multiplying the matrix in the left-hand side of the above inequality by $\Id\oplus W^{-1}\oplus \Id$,  one gets, for all $\tau\in\T$
$$
\scalebox{0.85}{$
\begin{bmatrix}
\He((A+BK)W)&-BK&BS-WK\tr-Z\tr\\
\star&-Q(\tau)&K\tr\\
\star&\star&-2S
\end{bmatrix}\prec 0$}
$$
which reads as \eqref{eq:thm:LMI_LES} with $J=0$. Therefore, since \eqref{eq:lemm:PolyInclusion} and \eqref{eq:thm:LMI_LES} hold, by invoking Theorem~\ref{theo:ExpStab} the result is proven. \QEDA
\end{pf}
\begin{remark}
Compared to Theorem~\ref{theo:ExpStab}, the estimate of the basin of attraction provided by Proposition~\ref{prop:ExpStabControl} is in general more conservative. This is due to enforcing $J=0$ and using inequality \eqref{eq:BoundW}. Therefore, once a control gain $K$ is designed via Proposition~\ref{prop:ExpStabControl}, an additional analysis stage based on Proposition~\ref{prop:FiniteNMI} can be performed to get a less conservative estimate of the basin of attraction. This aspect is discussed in Section~\ref{sec:Ex}.
 \hfill$\circ$
\label{rem:Analysis}
\end{remark}

\subsection{Optimization Aspects}
\label{sec:Opti}
%\subsection{Optimization}
Similarly as in \cite{hu2006stability,tarbouriech2011stability}, with the objective of 
enlarging the basin of attraction of the set $\A$, we turn our control design problem into the following optimization problem:
\begin{problem}
\label{prob:Opti}
Design $K$ to ``maximize'' the size of $\mathcal{Q}_{\bar{\mu}}$.\hfill$\diamond$ 
\end{problem}
To achieve this goal, one of the main difficulties that needs to be tackled is that the estimate of the basin of attraction provided by the set $\mathcal{Q}_{\bar{\mu}}$ defined in \eqref{eq:E_rho}
 is a subset of the set $\C$ defined in \eqref{eq:F}. In other words, this estimate is intrinsically dependent on the ``artificial'' variable $\tau$. This makes it difficult to define a convenient size criterion to enlarge the basin of attraction in the ``$(\xp, \xctilde)$-direction''. To overcome this drawback, next we provide a subset of $\mathcal{Q}_{\bar{\mu}}$ for which a convenient size criterion can be formulated.
\begin{lemma}
Let $\bar{\mu}>0$, $W\in\S^{\np}_+$, $R_1\in\S^{n_1}_+, R_2\in\S^{n_2}_+,\dots,R_{\ns}\in\S^{n_{\ns}}_+$, $S\in\Dy^{\nuu}_+$, $\sigma_1, \sigma_2,\dots, \sigma_{\ns}, \bar{\mu}\in\R_{>0}$ be given, $\T\ni\tau\mapsto\widehat{P}(\tau)$ be defined as in \eqref{eq:lemma:Phat}, and $\tau^\star\coloneqq (T_2^{(1)},T_2^{(2)},\dots, T_2^{(\ns)})$. Then, $
\mathcal{E}(\widehat{P}(\tau^\star),\bar{\mu})\times\T\subset\{x\in\C\colon \xcl\tr\widehat{P}(\tau)\xcl\leq\bar{\mu}\}
$.
\label{lemm:InnerBasin}
\end{lemma}
\smallskip
\begin{pf}
The proof follows directly from the fact that for each $\tau\in\T$, one has $\widehat{P}(\tau^\star)\succeq \widehat{P}(\tau)$. \QEDA
\end{pf}

Leveraging Lemma~\ref{lemm:InnerBasin}, Problem~\ref{prob:Opti} can be precisely formulated by selecting a suitable size criterion for the ellipsoidal set $\mathcal{E}(\widehat{P}(\tau^\star),\bar{\mu})$.
To this aim, different criteria are proposed in the literature; see \cite{Boyd}. In this work, we consider as a size criterion:
\begin{equation}
\Tr\left(\frac{\widehat{P}(\tau^\star)}{\bar{\mu}}\right)\!\!=\!\!\frac{\left(\Tr(W^{-1})+\sum_{i=1}^{\ns} \Tr(R_i) e^{\sigma_i T_2^{(i)}}\!\right)}{\bar{\mu}}.
\label{eq:Objective}
\end{equation}
In particular, larger values of $\Tr(\frac{1}{\bar{\mu}}\widehat{P}(\tau^\star))$ are associated to a smaller ellipsoid. To avoid dealing with $W^{-1}$ as a decision variable, we consider the following additional constraint
\begin{equation}
\label{eq:add:constraint}
\begin{bmatrix}
M_W&\Id_n\\
\star&W
\end{bmatrix}\succeq 0
\end{equation}
which is equivalent to $M_W-W^{-1}\succeq 0$. Hence, by setting $\bar{\iota}=\bar{\mu}^{-1}$, one can replace \eqref{eq:Objective} by:
$$
\bar{\iota}\left(\Tr(M_W)+\sum_{i=1}^{\ns} \Tr(R_i) e^{\sigma_i T_2^{(i)}}\right).
$$
%Pursuing this approach,  the following instance for 
Problem~\ref{prob:Opti}  can be solved by considering the following optimization scheme:
%can be considered:
\begin{equation}
\label{eq:OptiLMI}
\begin{aligned}
&\left\{
\begin{aligned}
&\minimize\scalebox{0.9}{$\varrho_1\bar{\iota}+\varrho_2\left(\Tr(M_W)+\sum_{i=1}^{\ns} \Tr(R_i) e^{\sigma_i T_2^{(i)}}\right)$}\\
&\text{s.t.}\\
&W\in\Spn, \bigoplus_{j=1}^{\ns} R_j\in\Spn, \eqref{eq:prop:M_ConvexHull_LES}, \eqref{eq:prop:PolyInclusion_Design}, \eqref{eq:add:constraint}
\end{aligned}\right.
\end{aligned}
\end{equation}
where $\varrho_1,\varrho_2\in\R_{>0}$ are tuning parameters.

When the real numbers $\alpha, \sigma_1, \sigma_2, \dots, \sigma_{\ns}$ are selected, optimization problem (\ref{eq:OptiLMI}) is a semidefinite program, \ie, an optimization problem with linear objective and LMI constraints. Therefore, the solution to this problem can be efficiently obtained via numerical available software; see \cite{Boyd}. 
On the other hand, the nonlinearities involving the real numbers $\alpha$, $\sigma_i, i=1,2,\dots, \ns$ are easily manageable (at least when $\ns$ is small enough) in a numerical scheme by performing a grid search.   
\begin{remark}
For analysis purposes, an optimization problem wholly similar to \eqref{eq:OptiLMI} can be formulated using the conditions in  Proposition~\ref{prop:FiniteNMI} to get an estimate of the basin of attraction for a given control gain.\hfill$\circ$
\label{rem:OptiAnalysis}
\end{remark}

\section{Numerical Example}
\label{sec:Ex}
\begin{figure}
\psfrag{x1}[][][1]{$x_{p1}$}
\psfrag{x2}[][][1]{$x_{p2}$}
\includegraphics[width=\columnwidth]{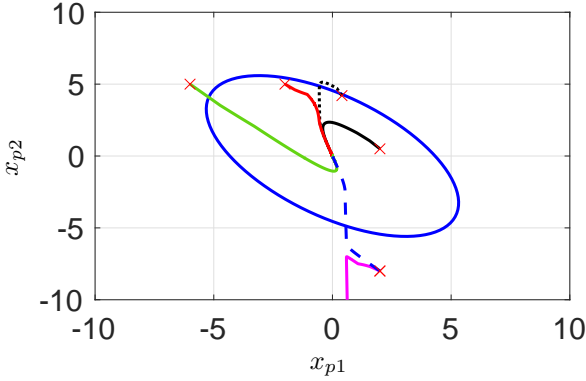}
\caption{Set $\mathcal{B}_{xp}$ (blue ellipse) and the response of the plant state from different initial conditions: $(0.4, 4.2, 0,0, T_2^{(1)}, T_2^{(2)})$ (dash black), $(2.5, -2.5, 0, 0, T_2^{(1)}, T_2^{(2)})$ (solid black), $(-8, 2, 0, 0, T_2^{(1)}, T_2^{(2)})$ (magenta), $(-6, 5, 0, 0, T_2^{(1)}, T_2^{(2)})$ (green), and $(-2, 5, 0, 0, T_2^{(1)}, T_2^{(2)})$ (red), and $(2, -8, -1.8, -5.7, T_2^{(1)}, T_2^{(2)})$ (dashed-blue line).}
\label{fig:Basin}
\end{figure}

In this section we showcase the effectiveness of the methodology proposed in this paper in an example\footnote{Numerical solutions to LMIs are obtained through the solver \textit{SDPT3} \cite{tutuncu2003solvingSDPT3} and coded in Matlab$^{\tiny{\textregistered}}$ via \textit{YALMIP} \cite{lofberg2004yalmip}. Simulations of hybrid systems are performed in Matlab$^{\tiny{\textregistered}}$ via the \textit{Hybrid Equations (HyEQ) Toolbox} \cite{sanfelice2013toolbox}.}.
Consider the following data for system \eqref{eq:Plant}:
$A=\begin{bmatrix}
-0.8 & -0.01\\ 1 & 0.1
\end{bmatrix}$, $B=\begin{bmatrix}
0.4\\ 0.1\end{bmatrix}$, $\bar{u}=1$, and 
$(T_2^{(1)}, T_2^{(2)})=(0.3, 0.7)$. As a first step, we design the control gain $K$ by solving optimization problem \eqref{eq:OptiLMI}, with $\varrho_1=\varrho_2=1$, via a line search on the parameters $\sigma_1, \sigma_2$, and $\alpha$. In particular, by selecting $\sigma_1=1.8$,  $\sigma_2=2.3$, and $\alpha=0.4$, one gets $K=\begin{bmatrix} -0.444 & -0.495\end{bmatrix}$.
As a second step, as suggested by Remarks~\ref{rem:Analysis}~and~\ref{rem:OptiAnalysis}, to reduce the conservatism, with the help of Lemma~\ref{lemm:Tech}, we solve an optimization problem wholly similar to \eqref{eq:OptiLMI} based on the conditions in Theorem~\ref{theo:ExpStab}. Following this approach, by selecting $\sigma_1=3.8
$ and $\sigma_2=2.3$, one gets
$W^{-1}=\begin{bmatrix}
0.0983 & 0.0788\\ 0.0788 & 0.0694 
\end{bmatrix}, R_1=0.0141, R_2=0.01172$, and $\bar{\mu}=2.66$.
To better assess the conservatism in the estimation of the basin of attraction, one can analyze the following set of plant initial conditions:  
\begin{equation}
\mathcal{B}_{x_{p}}\coloneqq\left\{\xp\in\R^{\np}\colon \xp\tr N\xp\leq \bar{\mu}\right\}
\label{eq:Bxp}
\end{equation}
where $N\coloneqq W^{-1}+(R_1e^{\sigma_1 T^{(1)}_{2}}\oplus R_2e^{\sigma_2 T^{(2)}_2})$. Notice that $\mathcal{O}\coloneqq\{(\xp, \xctilde)\in\R^{2\np}\colon \xp=\xctilde,  \xp\in\mathcal{B}_{x_{p}}\}\subset\mathcal{E}(P(\tau^\star),\bar{\mu})$, hence $\mathcal{O}$ is included in the basin of attraction of $\A$. Loosely speaking, the set $\mathcal{B}_{x_{p}}$ contains all plant initial conditions such that when the controller is initialized to zero, the closed-loop state converges to $\A$. \figurename~\ref{fig:Basin} depicts $\mathcal{B}_{\xp}$, along with some trajectories of the plant state from different initial conditions. In these simulations, jumps times are selected accordingly to a sinusoidal law with frequency $10$.
\begin{figure}
\psfrag{t}[][][1]{$t$}
\includegraphics[width=\columnwidth]{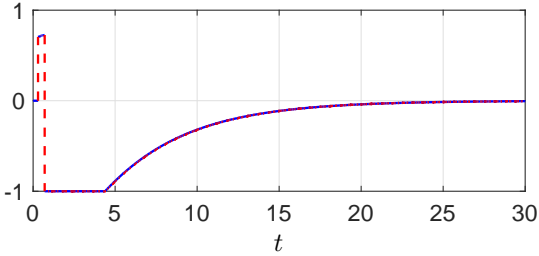}
\caption{Evolution of the control input from $x(0, 0)=(-2, 5, 0, 0, T_2^{(1)}, T_2^{(2)})$.}
\label{fig:Control}
\end{figure}
The first aspect is that the plant state may leave the set $\mathcal{B}_{x_{p}}$ and eventually converge to zero. This is testified by the trajectory from $x(0, 0)=(0.4, 4.2, 0, 0, T_2^{(1)}, T_2^{(2)})$ (dashed-black line). Another interesting aspect is illustrated by the trajectories from $x(0, 0)=\xi_1\coloneqq(2, -8, 0, 0, T_2^{(1)}, T_2^{(2)})$ (magenta line) and $x(0, 0)=\xi_2\coloneqq(2, -8, -1.8, -5.7, T_2^{(1)}, T_2^{(2)})$ (dashed-blue line). Although the two trajectories share the same initial conditions for the plant state, $\xi_1$ does not belong to the estimate of the basin of attraction, i.e., $\mathcal{E}(P(\tau^\star), \bar{\mu})\times\T$, while $\xi_2$ does, despite the fact that $(2, -8)\notin\mathcal{B}_{\xp}$. This explains why the trajectory from $\xi_2$ converges to the origin, whereas the one from $\xi_1$ does not. In \figurename~\ref{fig:Control}, we report the evolution of the control input from $x(0, 0)=(-2, 5, 0, 0, T_2^{(1)}, T_2^{(2)})$. The figure emphasizes that, as foreseen through our methodology, despite control input saturation, as illustrated in \figurename~\ref{fig:Basin} (red line), the plant state converges to zero. 
\section{Conclusion}
Feedback control design for linear systems with asynchronous aperiodic sampled state measurements and actuator saturation was studied. The proposed controller contains a copy of the plant whose state is updated whenever a new measurement is available, hence performing a fusion of data obtained at different time instances. 
The approach we pursued relies on Lyapunov theory for hybrid systems and the use of quadratic clock-dependent Lyapunov functions. Sufficient conditions in the form of matrix inequalities are given to ensure regional exponential stability of a set containing the origin of the plant. Despite the difficulty inherent to the clock-dependence of the Lyapunov function used, a numerical procedure, based on semidefinite programming techniques allowed to design the control gain and to characterize an inner estimate of the basin of attraction of $\A$ in the form of ellipsoidal sets.
Future research directions include the extension of the proposed methodology to the design of output feedback controllers in the presence of multi-rate output sampling and input saturation, for example by relying on the ideas presented in \cite{ferrante2018mathcal,merco2020lmi} and to the derivation of less conservative stability conditions. 
\bibliographystyle{plain}
\balance
\bibliography{biblio}

%%%%%%%%%%%%%%%%
\appendix
\section{Technical and ancillary results}
\begin{lemma}
\label{lemma:tj}
Let $\lambda_t>0$ and $\tau_D\coloneqq\min\{T_1^{(i)}, i\in\N\}$. Pick any $\lambda\in\left(0,\frac{\lambda_t \tau_D}{1+\tau_D}\right]$ and $\vartheta\geq\lambda \ns$.
Then, each solution $\phi$ to \eqref{eq:HybridCL} satisfies 
\begin{equation}
\label{eq:boundTJ}
-\lambda_t t\leq \vartheta-\lambda(t+j)
\end{equation}
for every $(t,j)\in\dom\phi$.
\end{lemma}
\begin{pf}
From \eqref{eq:boundTJ}, by rearranging the terms, one gets 
\begin{equation}
(-\lambda_t+\lambda)t+\lambda j-\vartheta\leq0
\label{eq:BoundTJ2}
\end{equation}
Now, pick any solution $\phi$ to hybrid system \eqref{eq:HybridCL}. From \eqref{eq:TimeBound}, it follows that for every $(t,j)\in\dom\phi$ 
\begin{equation}
\label{eq:BoundJ}
j\leq \frac{t}{\tau_D}+\ns.
\end{equation}
Then, for every strictly positive real number $\lambda$, from  the latter expression, and for every $(t,j)\in\dom\phi$, one gets
\begin{equation}
(-\lambda_t+\lambda)t+\lambda j-\vartheta\leq \left(-\lambda_t+\lambda+\frac{\lambda}{\tau_D}\right)t+\lambda \ns-\vartheta.
\end{equation}
Thus, being $\tau_D$ strictly  positive, by selecting 
$$\lambda\in\left(0,\frac{\lambda_t \tau_D}{1+\tau_D}\right], \vartheta\geq  \ns\lambda$$
yields \eqref{eq:BoundTJ2}, which concludes the proof.\QEDA
\end{pf}
\begin{property}
\label{prop:Lyapunov}
Let $\mathcal{T}\coloneqq \bigtimes_{i=1}^{\ns} [0, T_2^{(i)}]$, and $\C$, $f$, $\D$, and $G$ be defined as in \eqref{eq:F} and \eqref{eq:G}. There exist a continuously differentiable function $V\colon\R^{2n+\ns}\rightarrow\R$, positive real numbers $k, c_1, c_2$, and $c_3$, and a neighborhood $\mathcal{U}\subset\R^{2n}$ of $\{0\}$ such that 
\begin{align}
\label{eq:Sand}
&c_1 \vert x\vert^{k}_{\mathcal{A}}\leq V(x)\leq c_2 \vert x\vert^{k}_{\mathcal{A}}&\forall x\in \C,\\
\label{eq:FlowLyap}
&\langle\nabla V(x), f(x)\rangle\leq -c_3V(x)&\forall x\in \U\times\T,\\
\label{eq:JumpLyap}
&V(g)-V(x)\leq 0 &\forall x\in \D, g\in G(x).
\end{align}
$\hfill\triangle$
\end{property}
\begin{theorem}
\label{theo:LES}
Let Property~\ref{prop:Lyapunov} hold. Then, the set $\mathcal{A}$ defined in \eqref{eq:P2:Chap3:A} is exponentially stable for \eqref{eq:HybridCL}. Moreover, the basin of attraction of $\mathcal{A}$ contains any sublevel set of $\left.V\right\vert_{\dom V\cap\C}$ included in $\mathcal{U}\times\T$. 
\end{theorem}
\begin{pf} 
First of all, observe that all maximal solutions to \eqref{eq:HybridCL} are complete. 
Pick $\mu>0$ such that $L_V(\mu)\cap\C\subset \U\times\T$. Notice such a constant $\mu$ exists due to $\U$ being a neighborhood of the origin and \eqref{eq:Sand}. In particular, $L_V(\mu)\cap\C$ contains a neighborhood of $\A$.
Observe that the satisfaction of \eqref{eq:FlowLyap} and \eqref{eq:JumpLyap} ensures that $L_V(\mu)$ is strongly forward invariant\footnote{A set $S$ is strongly forward invariant for \eqref{eq:HybridCL} if any maximal solution $\phi$ to \eqref{eq:HybridCL} from $S$ is complete and $\range\phi\subset S$.} for \eqref{eq:HybridCL}. 
Pick any maximal solution $\phi$ to \eqref{eq:HybridCL} from $L_V(\mu)\cap\C$. 
Then, since $\range\phi\subset L_V(\mu)\cap\C$, thanks to \eqref{eq:FlowLyap}, direct integration of $(t, j)\mapsto V(\phi(t, j))$ yields 
$$
V(\phi(t, j))\leq e^{-c_3 t} V(\phi(0, 0))\quad \forall (t, j)\in\dom\phi,
$$
which, by using \eqref{eq:Sand}, gives:
$$
\vert\phi(t, j)\vert_{\mathcal{A}}\leq e^{-t\frac{c_3}{k}} \left(\frac{c_2}{c_1}\right)^{\frac{1}{k}}\vert\phi(0, 0)\vert_{\mathcal{A}}\quad \forall (t, j)\in\dom\phi.
$$
To conclude, by using Lemma~\ref{lemma:tj}, from the above relation, one gets 
$$
\vert\phi(t, j)\vert_{\mathcal{A}}\leq e^{-\lambda(t+j)+\vartheta}  \left(\frac{c_2}{c_1}\right)^{\frac{1}{k}}\vert\phi(0, 0)\vert_{\mathcal{A}}\quad \forall (t,j)\in\dom\phi,
$$
for some (solution independent) $\lambda, \vartheta>0$. This concludes the proof.\QEDA
\end{pf}
\begin{lemma}
\label{lemm:V}
Let, for $i=1, 2,\dots, \ns$, $U_i\colon\R^{n_{p_{i}}}\times\R_{\geq 0}\rightarrow\R$ be positive semidefinite with respect to $\{0\}\times[T_1^{(i)}, T_2^{(i}]$ and $V_1\colon\R^{n}\rightarrow\R$. Define for each $x\in\R^{\nx}$: 
$$
V(x)\coloneqq V_1(\xp)+\sum_{i\in\Q}U_i(\tildexci,\tau_i).
$$ 	
Then, the following inequality holds:
\begin{equation}
\label{eq:jumpClosedLoop}
V(g)-V(x)\leq 0\qquad\forall x\in\D, g\in G(x).
\end{equation}
\end{lemma}
\medskip
\begin{pf}
Pick $x=(\xp,\tilde{\xc},\tau)\in\D$ and $g\in G(x)$ and observe that by definition of $G$, there exists a unique $i\in\Q$, such that $g=(\xp,\gxc_i(\tilde{\eta}),v^{(i)})$, where $v^{(i)}=(v^{(i)}_1,v^{(i)}_2,\dots,v^{(i)}_{\ns})\in \Gt_i(\tau)$ with $\Gt_i$ defined as in \eqref{eq:Ti}. Assume without any loss of generality that $i=1$. Then, 
$$
\begin{aligned}
&\gxc_1(\tilde{\eta})=(0_{n_{p_1}}, \eta_{n_{p_2}},\dots,\eta_{n_{p_{\ns}}})\\
&\Gt_1(\tau)=[T_1^{(1)}, T_2^{(1)}]\times \{(\tau_2,\tau_3,\dots,\tau_{\ns})\},
\end{aligned}
$$
which gives $V(g)-V(x)=-U_1(\tilde{\eta}^{(1)},v^{(1)}_1)\leq 0$, $U_1$ being positive semidefinite with respect to $\{0\}\times[T^{(1)}_1, T^{(1)}_2]$. This concludes the proof.\QEDA	
\end{pf}

\begin{lemma}
	\label{lemm:ConvLin}
	Let $V$ and $W$ be two linear  real vector spaces, respectively, of dimension $n$ and $p$. Let $\mathscr{L}\colon V\rightarrow W$ be a linear application and $U\subset V$. Then, the following identity holds
	$$
	\mathscr{L}(\Co U)=\Co\mathcal{L}(U).
	$$ 
\end{lemma}
\begin{pf}
	We show the assert in two steps. First, we show that $\mathscr{L}(\Co U)\subset\Co\mathscr{L}(U)$. Pick $w\in\mathscr{L}(\Co U)$ and $v\in\Co U$ such that $w=\mathscr{L}(v)$.
	Then, since by assumption $U$ is isomorphic to $\R^n$, by the Carath\'eodory theorem, it follows that there exist some nonnegative real numbers $\gamma_1,\gamma_2,\dots,\gamma_{n+1}$, with $\sum_{i=1}^{n+1}\gamma_i=1$, and some points $v_1,v_2,\dots, v_{n+1}\in U$ such that $w=\mathscr{L}(\sum_{i=1}^{n+1}\gamma_i v_i)$. Therefore, since for all 
$i=1, 2,\dots, n+1$, $v_i\in U$, one has that $\mathscr{L}(v_i)\in \mathscr{L}(U)$, which, by linearity of $\mathscr{L}$, in turn implies $$
	w=\mathscr{L}\left(\sum_{i=1}^{n+1}\gamma_i v_i\right)=\sum_{i=1}^{n+1}\gamma_i \mathscr{L}(v_i)\in \Co\mathscr{L}(U).
	$$
Namely $\mathscr{L}(\Co U)\subset\Co\mathscr{L}(U)$. Now we show the complementary inclusion. Pick $w\in\Co\mathscr{L}(U)\subset W$. Due to $W$ isomorphic to $\R^p$, there exist
nonnegative real numbers $\beta_1,\beta_2,\dots,\beta_{p+1}$ with $\sum_{i=1}^{p+1}\beta_i=1$ and points $w_1,w_2,\dots, w_{p+1}\in\mathscr{L}(U)$ such that $w=\sum_{i=1}^{p+1}\beta_i w_i$.
For each $i=1,2,\dots, p+1$, pick $v_i\in U$ such that $w_i=\mathscr{L}(v_i)$; such points exist since $w_i\in\mathscr{L}(U)$ for each $i=1,2,\dots, p+1$. Then from the above relation, one gets
$w=\sum_{i=1}^{p+1}\beta_i w_i=\sum_{i=1}^{p+1}\beta_i \mathscr{L}(v_i)$.
Thus, still by linearity of $\mathscr{L}$, one gets $w=\mathscr{L}\left(\sum_{i=1}^{p+1}\beta_i v_i\right)\in\mathscr{L}(\Co U)$ and this concludes the proof.\QEDA
\end{pf}
\end{document}